\documentclass[twocolumn,showpacs,preprintnumbers]{revtex4}
\usepackage{graphicx}
\usepackage{dcolumn}
\usepackage{bm}
\usepackage{subfigure}

\begin{document}
\newcommand{\ds}{\displaystyle}
\newcommand{\be}{\begin{equation}}
\newcommand{\en}{\end{equation}}
\newcommand{\bea}{\begin{eqnarray}}
\newcommand{\ena}{\end{eqnarray}}
\title{Canonical and phantom scalar fields as an interaction of two perfect fluids}
\author{Mauricio Cataldo}
\altaffiliation{mcataldo@ubiobio.cl} \affiliation{Departamento de
F\'\i sica, Facultad de Ciencias, Universidad del B\'\i o-B\'\i o,
Avenida Collao 1202, Casilla 5-C, Concepci\'on, Chile.}
\author{Fabiola Ar\'evalo}
\altaffiliation{farevalo@udec.cl} \affiliation{Departamento de F\'{\i}sica, Universidad de Concepci\'{o}n,\\
Casilla 160-C, Concepci\'{o}n, Chile.}
\author{Patricio Mella}
\altaffiliation{patriciomella@udec.cl} \affiliation{Departamento de F\'{\i}sica, Universidad de Concepci\'{o}n,\\
Casilla 160-C, Concepci\'{o}n, Chile.}\date{\today}

\begin{abstract}
In this article we investigate and develop
specific aspects of Friedmann-Robertson-Walker (FRW) scalar field
cosmologies related to the interpretation that canonical and phantom
scalar field sources may be interpreted as cosmological
configurations with a mixture of two interacting barotropic perfect
fluids: a matter component $\rho_1(t)$ with a stiff equation of
state ($p_1=\rho_1$), and an ``effective vacuum energy" $\rho_2(t)$
with a cosmological constant equation of state ($p_2=-\rho_2$). An
important characteristic of this alternative equivalent formulation
in the framework of interacting cosmologies is that it gives, by
choosing a suitable form of the interacting term $Q$, an approach
for obtaining exact and numerical solutions. The choice of $Q$
merely determines a specific scalar field with its potential, thus
allowing to generate closed, open and flat FRW scalar field
cosmologies.
\end{abstract}
\smallskip
\maketitle 



\section{Introduction}
It is well known that scalar fields play an important role in
modeling both the early and late stages in the evolution of the
Universe
\cite{Copeland,Copeland1,Copeland2,Copeland3,Copeland4,Copeland5,
Copeland6,Copeland7,Copeland8,Copeland9,Copeland10,Copeland11}.
Inflation is considered an essential part of the description of the
early universe
\cite{Linde,Linde1,Linde2,Linde3,Linde4,Linde5,Linde6}. It does
solve classical cosmological problems such as flatness, horizon and
monopole problems, providing at the same time precise predictions
for the primordial density inhomogeneities, which are in good
agreement with observational data \cite{Hinshaw}. Many different
physical fields and mechanisms may be responsible for this
accelerating cosmic expansion
\cite{Felippe,Felippe2,Felippe3,Felippe4,Felippe5,Felippe6,Felippe7,
Felippe8,Felippe9,Felippe10,Felippe11,Felippe12,Felippe13,Felippe14,
Felippe15,Felippe16,Felippe17,Copeland4}, however, it has become a
common practice to employ scalar fields in order to explain this
early accelerated expansion \cite{Bamba15}. On the other hand, in
recent years, several proposals have been made, by using scalar
fields, in order to explain the late acceleration of the Universe
\cite{Elizalde} and the coincidence problem. In general, due to
isotropy and homogeneity of the Universe on cosmological scales,
most of the papers consider self-interacting scalar fields in the
FRW framework. We shall consider here canonical and phantom scalar
fields in the FRW cosmology described by the metric
\begin{eqnarray}
ds^2=dt^2 -a(t)^2 \left(\frac{dr^2}{1-kr^2}+r^2(d \theta^2 +\sin^2
\theta d \varphi^2) \right),
\end{eqnarray}
minimally coupled to gravity. The evolution of a scalar field
cosmology is given by
\begin{eqnarray}\label{FE}
3 H^2+\frac{3k}{a^2} &=& \kappa \rho_{\phi}, \\ 
\epsilon \dot{\phi} \ddot{\phi}+3 \epsilon H \dot{\phi}^2 &=&
\dot{\phi} V^{\prime}, \label{CETOTAL}
\end{eqnarray}
where $H=\dot{a}/a$, $\kappa=8 \pi G$, $\phi(t)$ is the scalar
field, $V(\phi)$ its potential, $\cdot = d/dt$, $\prime =d/d \phi$
and $\epsilon =\pm 1$. The Eq.~(\ref{CETOTAL}) represents the
dynamic equation for the evolution of a scalar field in a FRW
universe. The energy density and pressure of a such scalar field are
defined by the relations
\begin{eqnarray}\label{rhodeSF}
\rho_{\phi}=\epsilon \dot{\phi}^2/2+V(\phi),
\,\,\,\,\, p_{\phi}=\epsilon \dot{\phi}^2/2-V(\phi), \label{pdeSF}
\end{eqnarray}
where $\epsilon=1$ and $\epsilon=-1$ correspond to canonical and
phantom scalar fields respectively.

The corresponding scalar field equation of state (EoS) is given by
\begin{eqnarray}
\omega_{_{\phi}}=\frac{\epsilon \dot{\phi}^2/2-V(\phi)}{\epsilon
\dot{\phi}^2/2+V(\phi)}=\frac{\dot{\phi}^2/2-\epsilon V(\phi)}{
\dot{\phi}^2/2+\epsilon V(\phi)}.
\end{eqnarray}
Note that $\omega_{_{\phi}}$ varies in the range $-1<
\omega_{_{\phi}} <1$ for $V>0$, $\epsilon=1$, and for $V<0$,
$\epsilon=-1$.

Notice that, in order to fulfill the isotropy and homogeneity
requirements of FRW models, scalar fields actually have the form
$\phi=\phi(t)$, $V=V(t)$, so $\rho_{\phi}=\rho_{\phi}(t)$,
$p_{\phi}=p_{\phi}(t)$ and $\omega_{_{\phi}}=\omega_{_{\phi}}(t)$
depend also only on the cosmological time $t$.

In this paper we show that, any self-interacting scalar field
coupled to a FRW cosmology may be equivalently formulated as a
mixture of two interacting perfect fluids.

Perfect fluid description for sources plays an important role in
cosmology, where are considered single fluid configurations, as well
as multifluid ones, for describing the expansion of the Universe. It
is well known that for simplicity, and because it is consistent with
much we have observed about the universe, usually some epochs of the
universe evolution are modeled with perfect fluids and also with
mixtures of non interacting perfect fluids. Nevertheless, there are
no observational data confirming that this is the only possible
scenario. This means that it is theoretically plausible to consider
cosmological models containing perfect fluids which interact with
each other~\cite{Chattopadhyay1,Cataldo915}. This also applies to
the description of dark energy models for explaining the accelerated
expansion of the universe at the present stage.

In the case of a two-fluid FRW configuration, the cosmological
description of a superposition of two perfect fluids $\rho_1(t)$ and
$\rho_2(t)$, with barotropic EoS $p_1(t)=\omega_1 \rho_1(t)$ and
$p_2(t)=\omega_2 \rho_2(t)$, allows us to describe an effective
fluid $\rho_{_{eff}}(t)=\rho_1(t)+\rho_2(t)$ with an effective
pressure given by
\begin{eqnarray}
p_{_{eff}}(t)=\omega_{_{eff}}(t) \rho_{_{eff}}(t)=\frac{\omega_1
\rho_1(t)+\omega_2 \rho_2(t)}{\rho_1(t)+\rho_2(t)} \,
\rho_{_{eff}}(t).
\end{eqnarray}
Let us suppose that $\rho_1(t)>0$, $\rho_2(t)>0$ and $\omega_1>
\omega_2$, then the effective state parameter $\omega_{_{eff}}(t)$
varies in the range $\omega_2 < \omega_{_{eff}} < \omega_1$. Thus,
one can mimic the scalar field behavior with a two-fluid description
by requiring $\omega_1=1$ and $\omega_2=-1$.

With this purpose in mind we shall identify the first perfect fluid
$\rho_1$ with the scalar field $\phi$ as
\begin{eqnarray}\label{rho1}
\rho_{1}(t)=\epsilon \dot{\phi}^2/2,  \,\,\,\,\, p_{1}(t)=\epsilon
\dot{\phi}^2/2,
\end{eqnarray}
while the second fluid $\rho_2$ with the potential $V$ as
\begin{eqnarray}\label{rho2}
\rho_{2}(t)= V(\phi),  \,\,\,\,\, p_{2}(t)= -V(\phi).
\end{eqnarray}
This identification is based on the fact that both terms, $\epsilon
\dot{\phi}^2/2$ and $V$, contribute to the right hand side of
Einstein's equations, serving as the source of the gravitational
field. So, this interpretation means that the first matter component
is a perfect fluid with a stiff EoS, while the second component is
an ``effective vacuum energy" with a cosmological constant EoS.

If we require that each component, $\rho_1(t)$ and $\rho_2(t)$,
satisfies the standard conservation equation separately, then
$\rho_{eff}=C_1+C_2 a(t)^{-6}$ and the scale factor is given by
$a(t)=[-C_1^{-1}C_2 \cosh(\sqrt{3\kappa C_1}(t+C))]^{1/6}$. This
implies that we can not reproduce all the richness of scalar field
cosmologies. In order to do this, one must consider interacting FRW
models.

For two interacting fluids $\rho_1(t)$ and $\rho_2(t)$, with
barotropic EoS $\rho_1(t)= \omega_1 p_1(t)$ and $\rho_2(t)= \omega_2
p_2(t)$, we can write for the conservation equations
\begin{eqnarray}
\dot{\rho_1}+3H(1+\omega_1) \rho_1 &=& Q(t), \label{CE1}\\
\dot{\rho_2}+3H(1+\omega_2) \rho_2 &=& -Q(t), \label{CE2}
\end{eqnarray}
where $Q(t)$ is the interaction term. Note that for $Q>0$ we have a
transfer of energy from the fluid $\rho_2$ to fluid $\rho_1$, and
for $Q<0$ from the fluid $\rho_1$ to $\rho_2$.

By putting expressions~(\ref{rho1}) into Eq.~(\ref{CE1}),
and~(\ref{rho2}) into Eq.~(\ref{CE2}), with $\omega_1=1$ and
$\omega_2=-1$, we respectively obtain
\begin{eqnarray}
\epsilon \dot{\phi} \ddot{\phi}+3 \epsilon H \dot{\phi}^2 &=& Q(t),
\label{CE1A}
\\
\dot{\phi} V^{\prime} &=& -Q(t). \label{CE2A}
\end{eqnarray}
It is clear that Eqs.~(\ref{CE1A}) and~(\ref{CE2A}) directly imply
the fulfillment of the evolution equation of a scalar
field~(\ref{CETOTAL}). This shows that FRW scalar field models can
be seen as configurations with a mixture of a stiff perfect fluid
interacting with an ``effective vacuum energy". Note that for a
vanishing interaction term we have the condition $\dot{\phi}
V^{\prime}=0$. This implies that we are in presence of a
cosmological constant (for $\phi=const$ with $V=V(\phi)=const$), or
in the presence of a mixture of a stiff matter with a cosmological
constant (for $\phi=\phi(t)$ with $V=const$). Thus, clearly an
interaction term $Q \neq 0$ allows the cosmological constant
($p=-\rho=const$) to become a dynamical quantity, always respecting
the EoS $p(t)=-\rho(t)$.

Let us remark that to our knowledge this interpretation for scalar
fields, described in terms of two interacting fluids, was first
mentioned in the Ref.~\cite{Chimento15}, where the author finds the
group of symmetry transformation generated by interacting fluids in
spatially flat accelerated FRW cosmologies, and applies it to a
mixture of $n$ non-interacting canonical scalar fields with
exponential potential, in order to have assisted inflation. It must
be noted also that in the Ref.~\cite{Chimento15A} this
interpretation is applied to a flat FRW cosmology with a
self-interacting conformal scalar field $\Psi$. Specifically, it is
considered the potential $V(\Psi)=\lambda \Psi^4+V_0$ ($V_0>0$), and
it is shown that this scalar field may be represented as the
interaction of two perfect fluids with EoS $p_1= \rho_1 /3$ and
$p_2=-\rho_2=const$, representing radiation and vacuum energy
respectively. However, no one of this papers considers specific
interacting terms $Q$.

\section{Some applications}
Now, let us consider some exact cosmological scenarios containing a
scalar field $\phi$, described by a potential V. The question of the
choice of the interacting term remains crucial, so we would like to
make this choice motivated by considered in the literature specific
forms for $Q(t)$
\cite{Chattopadhyay,Chattopadhyay1,Chattopadhyay2,Chattopadhyay3,Chattopadhyay4,Chattopadhyay5}.
In the following, we shall obtain some known exact solutions for
self-interacting scalar field FRW cosmologies (with k=-1,0,1).

We shall begin by considering an interacting term given by
\begin{eqnarray}
Q(t)=3 \alpha H \rho_1.
\end{eqnarray}
where $\alpha$ is a constant parameter. In this case from
Eqs.~(\ref{CE1}) and~(\ref{CE2}) we obtain that for $\omega_1=1$ and
$\omega_2=-1$ the energy densities are given by
\begin{eqnarray}\label{rhoexp1}
\rho_1(a)&=&C_1 a^{3 (\alpha-2)},
\\
\rho_2(a)&=&C_2+\frac{C_1 \alpha}{2-\alpha} \, a^{3 ( \alpha-2)},
\,\,\,\,\,\,\,\,\,\, \label{rhoexp2}
\end{eqnarray}
and for $\omega_1=-1$ and $\omega_2=1$ the energy densities are
given by
\begin{eqnarray}\label{rhoexp15}
\rho_1(a)&=&C_1 a^{3 \alpha},
\\
\rho_2(a)&=&C_2 a^{-6}-\frac{C_1 \alpha}{\alpha+2} \, a^{3 \alpha}.
\,\,\,\,\,\,\,\,\,\, \label{rhoexp25}
\end{eqnarray}
Note that in both cases for $C_2=0$ we have that $\rho_1 \sim
\rho_2$. The interpretation is direct: for $\omega_1=1$ and
$\omega_2=-1$ the interacting term is proportional to the stiff
matter energy density ($\rho_s$), while for $\omega_1=-1$ and
$\omega_2=1$ the interacting term is proportional to effective
vacuum energy density ($\rho_v$).

In order to find the scale factor, we must use the Friedmann
equation in the form
\begin{eqnarray}\label{FEE}
3 H^2+\frac{3k}{a^2}=\kappa \rho_{1}(a)+\kappa \rho_{2}(a).
\end{eqnarray}
In general, we can not find the scale factor in analytical form, so
sometimes we shall look for some particular solutions to Friedmann
equation~(\ref{FEE}).

Now we shall discuss solutions for energy densities~(\ref{rhoexp1})
and~(\ref{rhoexp2}). Solutions for energy densities in the
form~(\ref{rhoexp15}) and~(\ref{rhoexp25}) will be discussed in Sec.
III.

\subsection{Exponential potential in flat FRW cosmologies}
First, let us consider the case $C_2=0$. Thus from
Eqs.~(\ref{rhoexp1}) and~(\ref{rhoexp2}) we have that
$r=\rho_2/\rho_1=\frac{\alpha}{2-\alpha}$. In this case, for $k=0$,
the scale factor is given by
\begin{eqnarray}
a(t)=a_0 (t+C)^{2/(6-3 \alpha)}.
\end{eqnarray}
Thus the energy densities~(\ref{rhoexp1}) and~(\ref{rhoexp2}) take
the form
\begin{eqnarray}\label{SFrhoexp1}
\rho_1(t)&=&\frac{\tilde{C}_1}{(t+C)^2},
\\
\rho_2(t)&=&\frac{\tilde{C_1} \alpha}{(2-\alpha)(t+C)^2},
\label{SFrhoexp2}
\end{eqnarray}
where $\tilde{C_1}=C_1 a_0^{-3(2-\alpha)}$. By comparing
Eqs.~(\ref{rho1}) and~(\ref{SFrhoexp1}) we find that $\epsilon
\dot{\phi}^2=\tilde{C}_1 (t+C)^{-2}$, implying that the scalar field
has the form
\begin{eqnarray} \label{phipotentialexp}
\phi(t)=\pm \sqrt{\epsilon \tilde{C}_1 }\ln (t+C)+\phi_0,
\end{eqnarray}
where $\phi_0$ is an integration constant. Now from
Eqs.~(\ref{rho2}) and~(\ref{SFrhoexp2}), and by taking into account
Eq.~(\ref{phipotentialexp}), we have that the potential is given by
\begin{eqnarray} \label{Vpotentialexp}
V(\phi)=\frac{\tilde{C}_1 \alpha}{(2-\alpha)}e^{\pm
(\phi(t)-\phi_0)/\sqrt{\epsilon\tilde{C}_1}}.
\end{eqnarray}
Note that for a canonical scalar field (i.e. $\epsilon=1$ and the
kinetic term of the scalar field is positive) we have that
$\tilde{C}_1>0$ and the energy density $\rho_1>0$. For $0<\alpha< 2$
we have a positive potential ($\rho_2>0$) and for $\alpha> 2$ or
$\alpha<0$ we have a negative scalar field potential ($\rho_2<0$).
In the framework of this perfect fluid interpretation, we conclude
that for $0<\alpha< 2$ (and $\alpha>2$) the interacting term $Q>0$,
implying that the energy is being transferred from the ``effective
vacuum energy" $\rho_2$ to the stiff fluid $\rho_1$, while for
$\alpha< 0$ we have that $Q<0$ and the energy goes from the stiff
fluid to the ``effective vacuum energy".

For a phantom scalar field (i.e. $\epsilon=-1$ and the kinetic term
of the scalar field is positive) we have that $\tilde{C}_1<0$ and
then Eq.~(\ref{SFrhoexp1}) implies that $\rho_1<0$. For $0<\alpha<
2$ we have a negative potential ($\rho_2<0$) and for $\alpha> 2$ or
$\alpha<0$ we have a positive scalar field potential ($\rho_2>0$).
On the other hand, in this case we have that for $0<\alpha< 2$ (and
$\alpha>2$) the interacting term $Q<0$ since $\rho_1<0$, implying
that the energy is being transferred from the stiff fluid $\rho_1$
to the ``effective vacuum energy" $\rho_2$, while for $\alpha< 0$ we
have that $Q>0$ and the energy goes from the ``effective vacuum
energy" to the stiff fluid.

It is remarkable that in this case the canonical and phantom scalar
fields have the same EoS, given by
\begin{eqnarray}
\omega_{\phi}=1-\alpha.
\end{eqnarray}
Thus we have that $-1<\omega_{\phi}<1$ for $0<\alpha<2$, while for
$\alpha>2$ and $\alpha<0$ it is verified that $\omega_{\phi}<-1$ and
$\omega_{\phi}>1$ respectively. In this case both types of scalar
fields cosmologies have the same scale factor.

It is interesting to note that we can put for energy
densities~(\ref{rhoexp1}) and~(\ref{rhoexp2}) $\omega_1=-1$ and
$\omega_2=1$ and identify $\rho_{1}(t)= V(\phi), \rho_{2}(t)=
\epsilon \dot{\phi}^2/2$, but in this case we can not write the
scale factor, for $C_1 \neq 0$ and $C_2 \neq 0$, in analytical form.
For the case $C_1 \neq 0$, $C_2=0$ the solutions are included
in~(\ref{phipotentialexp}) and~(\ref{Vpotentialexp}).

Notice that for energy densities~(\ref{rhoexp15})
and~(\ref{rhoexp25}), with $C_2=0$ we obtain also a scalar field
with an exponential potential.

\subsection{Sinusoidal potential in flat FRW cosmologies}
Let us now consider the energy densities~(\ref{rhoexp1})
and~(\ref{rhoexp2}) with $C_2\neq 0$. Thus from the Friedmann
equation, with $k=0$, we obtain that the scale factor is given by
\begin{eqnarray}
a(t)=a_0 \times  \,\,\,\,\,\,\,\,\,\,\,\,\,\,\,\,\,\,\,\,\,\,\,\,\,\,\,\,\,\,\,\,\,\,\,\,\,\,\,\,\,\,\,\,\,
\,\,\,\,\,\,\,\,\,\,\,\,\,\,\,\,\,\,\,\,\,\,\,\,\,\,\,\,\,\,\,\,\,\,\,\,\,\,\,\,\,\,\,\,\, \nonumber  \\
\left[\cosh \left(\sqrt{\frac{3 \kappa}{4} C_2}
(2-\alpha)(t+C)\right)\right]^\frac{2}{3(2-\alpha
)}, \nonumber \\
\end{eqnarray}
where $C$ is an integration constant, and the constraint $C_2
(\alpha-2)=2C_1 a_0^{3(\alpha-2)}$ must be fulfilled. Thus we can
write energy densities as functions of the cosmological time $t$
\begin{eqnarray}\label{tanhrhoexp1}
\rho_1(t)&=&\frac{C_2(\alpha-2)}{2 \cosh^2 \left(\sqrt{\frac{3
\kappa}{4} C_2} (2-\alpha)(t+C)\right)},   \\ \rho_2(t)&=&C_2 \times
\,\,\,\,\,\,\,\,\,\,\,\,\,\,\,\,\,\,\,\,\,\,\,\,\,\,\,\,\,\,\,
\,\,\,\,\,\,\,\,\,\,\,\,\,\,\,\,\,\,\,\,\,\,\,\,\,\,\,\,\,\,\,
\,\,\,\,\,\,\,\,\,\,\,\,\,\,\,\,\,\,\,\,\,\,\,\,\,\,\,\,\,\,\,
\,\,\,\,\,\,\,\,\,\,\,\,\,
\nonumber \\
&& \hspace{-0.5cm} \left[1-\frac{\alpha}{2}
\cosh^{-2} \left(\sqrt{\frac{3 \kappa}{4} C_2}
(2-\alpha)(t+C)\right) \right]. \,\,\,\,\,\,\,\,\,\,
\label{tanhrhoexp2}
\end{eqnarray}

Now we shall connect this interacting solution with a scalar field.
From Eqs.~(\ref{rho1}) and~(\ref{tanhrhoexp1}) we have that the
scalar field has the form
\begin{eqnarray}\label{phicaso2}
\phi(t)-\phi_0 &=& \frac{\pm 1}{(\alpha-2)}\sqrt{\frac{16 \epsilon
(\alpha-2)}{3\kappa }}\times \nonumber \\ &&  \arctan
\left(e^{\sqrt{3 \kappa C_2/4}(\alpha-2)(t+C)} \right),
\end{eqnarray}
while from Eqs.~(\ref{rho2}) and~(\ref{tanhrhoexp2}), and by taking
onto account Eq.~(\ref{phicaso2}), we find that the potential is
given by
\begin{eqnarray}
V(\phi)=C_2 \left[1-\frac{\alpha}{2} \sin^{2} \left(\sqrt{\frac{3
\kappa}{4 \epsilon (\alpha-2)}} (\alpha-2)(\phi-\phi_0)\right)
\right]. \nonumber \\
\end{eqnarray}
This type of potentials are considered for describing
natural/chaotic inflation~\cite{Alabidi}.

\subsection{Quadratic potential in non-flat FRW cosmologies}
Now we shall consider a solution for energy densities in the
form~(\ref{rhoexp1}) and~(\ref{rhoexp2}) with $k \neq 0$. This
implies that $\rho_1+\rho_2=\frac{2C_1}{2-\alpha}a^{-6+3
\alpha}+C_2$. Let us consider the particular solution $a(t)=a_0
e^{At}$, where $a_0$ and $A$ are constants. Thus in order to satisfy
the Friedmann equation~(\ref{FEE}) the following constraints must be
fulfilled: $\alpha=4/3$, $\kappa C_1=k$ and $\kappa C_2=3 A^2$, and
energy densities take the form
\begin{eqnarray}\label{kno0}
\kappa \rho_1=\frac{k}{a^2}, \,\,\,\,\,  \kappa
\rho_2=\frac{2k}{a^2}+3 A^2.
\end{eqnarray}
By comparing Eqs.~(\ref{rho1}) and~(\ref{rho2}) with
expressions~(\ref{kno0}) we find that
\begin{eqnarray}
\epsilon \dot{\phi}^2=\frac{2 k}{\kappa a^2}, \,\,\,\,\,
V=\frac{2k}{\kappa a^2}+\frac{3 A^2}{\kappa}.
\end{eqnarray}
Thus we have that the scalar field has the form
\begin{eqnarray} \label{SFk0}
\phi(t)=\phi_0 \pm \frac{\sqrt{2 \epsilon k/\kappa}}{A a_0} e^{-A
t},
\end{eqnarray}
where $\phi_0$ is an integration constant, while the potential is
given by
\begin{eqnarray} \label{potential1k0}
V(\phi)=\frac{3 A^2}{\kappa}+\epsilon A^2 (\phi-\phi_0)^2.
\end{eqnarray}
Note that for a canonical field we have that $\epsilon=k=1$ and then
$\rho_1>0$ and $\rho_2>0$, implying that the kinetic term and the
potential are always positive magnitudes. On the other hand, for a
phantom scalar field we have that $\epsilon=k=-1$ and then
$\rho_1<0$ and $\rho_2>0$ for $a> \sqrt{2 A^2/3}$, implying that the
kinetic term is always negative. In this case the interacting term
is given by $Q=4 H \rho_1$, thus for a canonical scalar field the
transfer of energy goes from the effective vacuum fluid to the stiff
one, and for a phantom scalar field the energy is being transferred
from the stiff fluid to the effective vacuum.

\subsection{Quadratic potential in flat FRW cosmologies}
Let us now consider an interacting term given by
\begin{eqnarray}\label{Qconst}
Q(t)=3 H \alpha,
\end{eqnarray}
where $\alpha$ is a constant parameter. For the state parameter
values $\omega_1=1$ and $\omega_2=-1$ the solution has the form
\begin{eqnarray}\label{30}
\rho_1(a)&=&\frac{\alpha}{2}+  C_1 a^{-6},
\\
\rho_2(a)&=&C_2 -3 \alpha \ln a. \label{31}
\end{eqnarray}
In order to find an explicit expression for the scale factor let us
consider the solution~(\ref{30}) and~(\ref{31}) with $C_1=0$. Thus
we find
\begin{eqnarray}
a(t)= a_0 e^{-\frac{\kappa \alpha}{4} \, (t+C)^2},
\end{eqnarray}
where $C$ is a constant of integration, and the energy densities
take the form
\begin{eqnarray}
\rho_1(t)&=& \frac{\alpha}{2}, \label{caso1}
\\
\rho_2(t)&=&\tilde{C}_2+\frac{3 \kappa \alpha^2}{4} \left(t+C
\right)^2,\label{caso2}
\end{eqnarray}
where $\tilde{C}_2=C_2 -3\alpha \ln a_0$. By comparing
Eqs.~(\ref{rho1}) and~(\ref{caso1}) we find that $\epsilon
\dot{\phi}^2=\alpha$, implying that the scalar field has the form
\begin{eqnarray} \label{potential1}
\phi(t)=\pm \sqrt{\epsilon \alpha}t+C_3,
\end{eqnarray}
where $C_3$ is an integration constant. Now from Eqs.~(\ref{rho2})
and~(\ref{caso2}) we have that the potential as function of the
cosmological time $t$ is given by $V(t)=\tilde{C}_2 +\frac{3 \kappa
\alpha^2}{4} \left(t+C \right)^2$. By rewriting the scalar
field~(\ref{potential1}) as $\phi(t)=\pm \sqrt{\epsilon
\alpha}(t+C)+\phi_0$ we can write the potential in the form
\begin{eqnarray}\label{Final Form potential1}
V(\phi)=\tilde{C}_2+\frac{3 \kappa \epsilon \alpha}{4}
(\phi-\phi_0)^2.
\end{eqnarray}
The Eq.~(\ref{Final Form potential1}) is one of the basic potentials
considered in standard cosmology.

Note that for $\omega_1=-1$ and $\omega_2=1$ the energy densities
are given by $\rho_1(a)=C_2 + 3 \alpha \ln a$ and $\rho_2(a)=
-\frac{\alpha}{2}+  C_1 a^{-6}$, implying that is the same solution
discussed above.

\subsection{Non linear interacting term}
Now we shall consider a non-linear coupling given by
\begin{eqnarray}\label{Qnolineal}
Q=3 \alpha H \frac{\rho_1 \rho_2}{\rho_1 +\rho_2}.
\end{eqnarray}
For the state parameter values $\omega_{1}= 1$ and $\omega_{2}=- 1$
the energy densities are given
by
\begin{eqnarray}\label{rho1nolineal}
\rho_1(a)&=&\frac{C_1}{C_2}
\, a^{3 \alpha-6} (C_1 a^{3
\alpha-6}+C_2)^{-\alpha/(\alpha-2)}, \\
\rho_2(a)&=& (C_1 a^{3 \alpha-6}+C_2)^{-\alpha/(\alpha-2)},
\label{rho2nolineal}
\end{eqnarray}
where $C_1$ and $C_2$ are constants of integration. Note that in
this case the energy densities satisfy the relation
$\rho_1/\rho_2=\frac{C_1}{C_2} \, a^{3 \alpha-6}$.

Now, in order to find an analytical solution we shall put
$\alpha=1$. Thus the scale factor is given by
\begin{eqnarray}
a(t)=\frac{C_2^{2/3}}{C_2} \left(e^{\pm \sqrt{3\kappa C_2}(t+C)}
-C_1 \right)^{1/3},
\end{eqnarray}
where $C$ is an integration constant. This implies that the energy
densities will be given by
\begin{eqnarray}
\rho_1(t)&=&\frac{C_1 C_2 e^{\pm \sqrt{3\kappa C_2}(t+C)}}{(e^{\pm
\sqrt{3\kappa C_2}(t+C)} -C_1)^2}, \\
\rho_2(t)&=&\frac{C_2 e^{\pm \sqrt{3\kappa C_2}(t+C)}}{e^{\pm
\sqrt{3\kappa C_2}(t+C)} -C_1}.
\end{eqnarray}
Let us now consider the positive branch, i.e.
$a(t)=\frac{C_2^{2/3}}{C_2} \left(e^{\sqrt{3\kappa C_2}(t+C)} -C_1
\right)^{1/3}$. Thus the scalar field and its potential have the
form
\begin{eqnarray}
\phi(t)-\phi_0=\sqrt{\frac{8\epsilon}{3 \kappa}} arctanh \left(
\frac{e^{
\frac{1}{2}\sqrt{3\kappa C_2}(t+C)}}{\sqrt{C_1}}\right), \\
V(\phi)=C_2 \left(1- \cosh^2 \left( \sqrt{\frac{3 \kappa}{8
\epsilon}} (\phi-\phi_0) \right) \right),
\end{eqnarray}
respectively.

Finally, note that due to the symmetry of the interacting
term~(\ref{Qnolineal}) with respect to energy densities $\rho_1$ and
$\rho_2$ the solution for $\omega_{1}=- 1$ and $\omega_{2}= 1$ is
given by $\rho_1(a)=(C_1 a^{-3 \alpha-6}+C_2)^{-\alpha/(\alpha+2)}$,
$\rho_2(a)= \frac{C_1}{C_2} \, a^{-3 \alpha-6} (C_1 a^{-3
\alpha-6}+C_2)^{-\alpha/(\alpha+2)}$, implying that is the same
solution of Eqs.~(\ref{rho1nolineal}) and~(\ref{rho2nolineal}).


\section{A procedure for finding the scalar field as function of
the scale factor $a$} Up to now we have shown how we can obtain
scalar field cosmologies by writing a scalar field as a function of
the cosmological time $t$. However one can make the same by using
the scale factor $a$ as a main variable. Let us suppose that the
energy densities $\rho_1$ and $\rho_2$ are given as functions of the
scale factor $a$. From Eq.~(\ref{rho1}) we obtain $\pm \sqrt{2
\epsilon \rho_1(a)}=\phi^{\prime}(a) \dot{a}$, while from the
Friedmann equation~(\ref{FE}) we can write $\dot{a}=\pm
\sqrt{(\kappa/3)(\rho_1+\rho_2)a^2-k}$, thus obtaining
\begin{eqnarray} \label{SFdea}
\phi(a)=\pm \int \sqrt{\frac{6 \epsilon
\rho_1(a)}{\kappa[\rho_1(a)+\rho_2(a)]a^2-3k}} \, da+\phi_0,
\end{eqnarray}
where $\phi_0$ is a constant of integration. In order to write the
potential as $V=V(\phi)$ we can write from~(\ref{SFdea}) the scale
factor as $a=a(\phi)$ and introducing it into Eq.~(\ref{rho2}) we
can write the potential as a function of the scale factor.

Note that we can find the scale factor as function of the
cosmological time with the help of the general expression
\begin{eqnarray}\label{generaladet}
t+C=\pm \int \, \frac{\phi^{\prime}}{\sqrt{2 \epsilon \rho_1(a)}} \,
da,
\end{eqnarray}
where $C$ is an integration constant.

\subsection{An application for $Q=3 H \alpha$}
As a simple example let us consider again the interacting scenario
$Q=3 H \alpha$. From Eqs.~(\ref{30}), (\ref{31}) and~(\ref{SFdea})
we obtain
\begin{eqnarray}\label{phideaexample}
\phi(a)-\phi_0=\pm \sqrt{\frac{2 \epsilon}{3 \kappa \alpha}\,
(\alpha+2 C_2-6 \alpha\ln a)},
\end{eqnarray}
where $\phi_0$ is an integration constant and we have put $C_1=0$.
Thus by taking into account Eq.~(\ref{rho2}) we have that the
potential is given by
\begin{eqnarray}
V(\phi)= -\frac{\alpha}{2}+\frac{3 \kappa \epsilon \alpha}{4}
(\phi-\phi_0)^2.
\end{eqnarray}
Now, from Eqs.~(\ref{generaladet}), (\ref{phideaexample}), and by
taking into account that $\rho_1(a)=\alpha/2$, we obtain that
\begin{eqnarray}
a(t) \sim e^{-\frac{\kappa \alpha}{4} \, (t+C)^{2}}.
\end{eqnarray}

\subsection{An application for $Q=3 \alpha H \rho_1$}
Now, we shall use our procedure to generate a scalar field cosmology
for interacting scenarios described by the coupling $Q=3 \alpha H
\rho_1$. In the following we shall found the general solution for
Eqs.~(\ref{rhoexp1}) and~(\ref{rhoexp2}) with $\alpha=4/3$. From
Eq.~(\ref{SFdea}) we find that
\begin{eqnarray}\label{phideaplicacion}
&& \phi(a)-\phi_0= \pm \sqrt{\frac{2 \epsilon C_1}{\kappa C_1-k}}
\times \nonumber \\
&& \ln \left( \frac{6(\kappa C_1-k)+2\sqrt{9(\kappa C_1-k)^2+3
\kappa C_2 (\kappa C_1-k)a^2})}{a}
\right) \nonumber \\
\end{eqnarray}
for $C_1 \neq k/\kappa$, and
\begin{eqnarray}\label{phi de apl}
\phi(a)-\phi_0= \pm \sqrt{\frac{6 \epsilon k}{\kappa^2  C_2 a^2}}
\end{eqnarray}
for $C_1= k/\kappa$.

Thus the potential is given by
\begin{eqnarray}
&& V(\phi) = C_2+\frac{C_1}{72 (\kappa C_1-k)^2} \times \nonumber \\
&&\left( e^{-\sqrt{\frac{\kappa C_1-k}{2C_1 \epsilon }} \phi}-12
\kappa C_2(\kappa C_1-k)e^{\sqrt{\frac{\kappa C_1-k}{2C_1 \epsilon
}} \phi}\right)^2
\end{eqnarray}
for $C_1 \neq k/\kappa$, and
\begin{eqnarray}\label{potencial particular}
V(\phi)=C_2+ \frac{\kappa \epsilon C_2}{3} \, (\phi-\phi_0)^2
\end{eqnarray}
for $C_1= k/\kappa$.

The scale factor may be found by using Eq.~(\ref{generaladet}) with
$\rho_1(a)=C_1/a^2$. Thus with the help of
Eq.~(\ref{phideaplicacion}) we have
\begin{eqnarray}
a(t)&=&\frac{1}{2\sqrt{\kappa C_2}} \times \nonumber \\ && \left(
e^{\sqrt{\frac{\kappa C_2}{3}} \, (t+C)} - 3 (\kappa C_1 -k)
e^{-\sqrt{\frac{\kappa
C_2}{3}} \, (t+C)} \right) \nonumber \\
\end{eqnarray}
for $C_1 \neq k/\kappa$, and with the help of Eq.~(\ref{phi de apl})
we find
\begin{eqnarray}\label{particular a}
a(t)=e^{\sqrt{\frac{\kappa C_2}{3}} \, (t+C)}
\end{eqnarray}
for $C_1= k/\kappa$.

From this solution we can obtain discussed before in the literature
scalar field cosmologies. For example the solution given by
Eqs.~(\ref{potencial particular}) and~(\ref{particular a}) is the
same solution given by Eqs.~(\ref{kno0})-(\ref{potential1k0}). They
are connected by the relation $\kappa C_2=3 A^2$.

On the other hand for $\kappa C_1=(k+1/3)$  the scale factor takes
the form $a(t)=(1/\kappa C_2)\sinh (\sqrt{\kappa C_2/3} \, (t+C))$

Lastly, let us consider a solution for energy densities in the
form~(\ref{rhoexp15}) and~(\ref{rhoexp25}). For $\alpha=-7/3$ from
eq.~(\ref{SFdea}) we have that $\phi(a)-\phi_0=\pm \int
\sqrt{\frac{6\epsilon C_1}{\kappa(-6 C_1 a^{2}+C_2 a^{3})}} \, da$,
obtaining for the scalar field
\begin{eqnarray}
\phi(a)-\phi_0= \pm 2 \sqrt{\frac{\epsilon}{\kappa}}  \arctan \,
\sqrt{\frac{-6 C_1+C_2 a}{6 C_1}}.
\end{eqnarray}
Thus by taking into account Eq.~(\ref{rhoexp15}) we conclude that
the potential has the form
\begin{eqnarray}
V(\phi)= \frac{C_1 C_2^7}{[6C_1(1 +\tan^2(\sqrt{\frac{\kappa}{4
\epsilon}} \, \phi))]^7}.
\end{eqnarray}
By using Eq.~(\ref{generaladet}) we find for the scale factor
\begin{eqnarray}
t+C=\pm \frac{\sqrt{a(C_2 a-6 C_1)}}{\sqrt{12 \kappa} C_2^3} \, (2
C_2^2 a^2+15C_1C_2 a+135 C_1^2)  \nonumber \\
\pm \frac{135 C_1^3}{4} \sqrt{\frac{3}{\kappa C_2^7}} \, \ln
\frac{[C_2 a-3 C_1+\sqrt{C_2 a(C_2 a-6 C_1)} \, ]^2}{C_2}. \nonumber
\\
\end{eqnarray}

\subsection{An application for $Q=3 \alpha H \frac{\rho_1 \rho_2}{\rho_1+\rho_2}$}
Let us now consider the non-linear interacting term $Q=3 \alpha H
\frac{\rho_1 \rho_2}{\rho_1+\rho_2}$. We shall found the general
solution for Eqs.~(\ref{rho1nolineal}) and~(\ref{rho2nolineal}) with
$\alpha=5/3$. From Eq.~(\ref{SFdea}) we find that
\begin{eqnarray}\label{phinolineal15}
\phi(a)-\phi_0=-2 \sqrt{\frac{6 \epsilon}{\kappa}} \, arctanh
\left(\sqrt{\frac{C_1+C_2 a}{C_1}}   \right),
\end{eqnarray}
where $\phi_0$ is an integration constant. Consequently, by taking
into account Eqs.~(\ref{rho2}) and~(\ref{rho2nolineal}) the
potential is given by
\begin{eqnarray}
V(\phi)=-C_2^5 \sinh^{10} \left(\sqrt{\frac{\kappa}{24 \epsilon}}
(\phi-\phi_0) \right).
\end{eqnarray}
Now, by using Eqs.~(\ref{rho1nolineal}), (\ref{generaladet})
and~(\ref{phinolineal15}) we find for the scale factor
\begin{eqnarray}
t+C=\sqrt{\frac{3}{4 \kappa C_2^5}} \times \nonumber \\
\frac{3 C_1^2+4 C_1 C_2 a + \ln \left[ \left(C_1+C_2 a
\right)^{2(C_1+C_2 \, a)^2} \right ]}{(C_1+C_2 \, a)^2}.
\end{eqnarray}
Lastly, as another example, note that by following the same
procedure we can directly rederive the solution for $\alpha=1$
discussed at the end of Sec. II.

\section{Some general relations for scalar fields}
Note that in the framework of FRW scalar field cosmologies we can
obtain some general relations involving $\phi$ and $V$ for specific
interacting terms $Q$. In general, in this alternative equivalent
formulation of scalar fields the barotropic perfect fluids $\rho_1$
and $\rho_2$ with EoS $\omega_{1,2}=\pm 1$ or $\omega_{1,2}=\mp 1$
must fulfill Eqs.~(\ref{CE1}) and~(\ref{CE2}). If the interaction
term has the form $Q=3 \alpha H \rho_1$ then the solution may be
written through $\rho_1(t)$ in the following form:
\begin{eqnarray}
\rho_2(t) &=& C_1
\rho_1(t)^{\frac{\omega_2+1}{\omega_1+1-\alpha}}+\frac{\alpha \,
\rho_1(t)}{\omega_1-\omega_2-\alpha}, \label{Gen1} \\
a(t) &=& C_2 \rho_1(t)^{-\frac{1}{\omega_1+1-\alpha}} \label{Gen2}.
\end{eqnarray}
This form has the advantage to provide us with a way to relate the
kinetic term $\dot{\phi}^2$ and scalar field potential $V$.

Let us first consider that the barotropic perfect fluids $\rho_1$
and $\rho_2$ are a stiff fluid $\rho_{s}$ and an effective vacuum
$\rho_{v}$, respectively. Therefore, by putting $\omega_1=1$ and
$\omega_2=-1$ into Eqs.~(\ref{Gen1}) we obtain
\begin{eqnarray}
\rho_2(t)=C_1+\frac{\alpha}{2-\alpha} \, \rho_1(t). \label{Gen1A}
\end{eqnarray}
Thus, by taking into account the Eqs.~(\ref{rho1}) and~(\ref{rho2})
we may rewrite Eq.~(\ref{Gen1A}) in the form
\begin{eqnarray}\label{Gen1AA}
V(t)=C_1+\frac{\alpha \epsilon}{2(2-\alpha)} \, \dot{\phi}^2.
\end{eqnarray}
On the other hand, if now the barotropic perfect fluids $\rho_1$ and
$\rho_2$ are respectively an effective vacuum fluid and a stiff one,
then by putting $\omega_1=-1$ and $\omega_2=1$ into
Eqs.~(\ref{Gen1}) we obtain
\begin{eqnarray}\label{Gen1B}
\rho_2(t)=C_1 \rho_1(t)^{-2/\alpha}-\frac{\alpha
\rho_1(t)}{2+\alpha}.
\end{eqnarray}
Now by taking into account the Eqs.~(\ref{rho1}) and~(\ref{rho2}) we
may rewrite Eq.~(\ref{Gen1B}) in the form
\begin{eqnarray}\label{Gen1BB}
\frac{\epsilon \dot{\phi}^2}{2}=C_1 V(t)^{-2/\alpha}-\frac{\alpha
V(t)}{2+\alpha}.
\end{eqnarray}
Thus any scalar field $\phi$, which satisfies
conditions~(\ref{Gen1AA}) or~(\ref{Gen1BB}) has an alternative
equivalent formulation in the framework of interacting cosmologies
with an interaction term of the form $Q=3 \alpha H \rho_{s}$ or $Q=3
\alpha H \rho_{v}$.

For interacting scenarios described by the interaction $Q=3 H
\alpha$ the solution may be written  through $\rho_1(t)$ in the form
\begin{eqnarray}\label{rho215A}
\rho_2(t)=C_2-\frac{\alpha}{2} \ln \left
(\frac{C_1}{\rho_1(t)-\alpha/2} \right), \\
a(t)=\left(\frac{C_1}{\rho_1(t)-\alpha/2} \right)^{1/6},
\end{eqnarray}
for $\omega_1=1$, $\omega_2=-1$ and $C_1 \neq 0$. Thus, by taking
into account the Eqs.~(\ref{rho1}) and~(\ref{rho2}) we may rewrite
Eq.~(\ref{rho215A}) in the form
\begin{eqnarray}
V(t)=C_2-\frac{\alpha}{2} \ln \left (\frac{2 C_1}{\epsilon
\dot{\phi}^2 -\alpha} \right).
\end{eqnarray}
Note that for the case $C_1=0$ the relation~(\ref{Final Form
potential1}) must be considered for $V$ and $\phi$.

For interacting scenarios described by $Q=3 \alpha H \frac{\rho_1
\rho_2}{\rho_1 +\rho_2}$ with $\omega_1=1$ and $\omega_2=-1$ the
solution may be written through $\rho_2(t)$ in the form
\begin{eqnarray}\label{nolinealrho1rho2}
\rho_1(t)&=& \left( \frac{\rho_2(t)^{(2-\alpha)/\alpha}-C_2}{C_2}
\right) \rho_2(t),
\\
a(t)&=&\left( \frac{\rho_2(t)^{(2-\alpha)/\alpha}-C_2}{C_1}
\right)^{1/(3 \alpha-6)}.
\end{eqnarray}
Thus, by taking into account the Eqs.~(\ref{rho1}) and~(\ref{rho2})
we may rewrite Eq.~(\ref{nolinealrho1rho2}) in the form
\begin{eqnarray}
\frac{\epsilon}{2} \, \dot{\phi}^2&=& \left(
\frac{V(t)^{(2-\alpha)/\alpha}-C_2}{C_2} \right) V(t).
\end{eqnarray}

\section{Conclusions}
Within the framework of FRW cosmologies scalar fields play a crucial
role in the description of accelerated expansion of the Universe.
This mainly because the scalar field state parameter varies in the
range $-1<\omega_{_{\phi}}<1$, inducing an accelerated expansion for
$\omega_{_{\phi}}<-1/3$. In this work we have shown that isotropic
and homogeneous scalar field cosmologies may be interpreted as
cosmological configurations with a mixture of a stiff perfect fluid
interacting with an ``effective vacuum energy". In the absence of an
interaction, we still have an accelerated expansion, due to the
presence of a cosmological constant ($\phi=const$ and
$V=V(\phi)=const$), or to the presence of a mixture of a stiff
matter with a cosmological constant ($\phi=\phi(t)$ and $V=const$).
Therefore the interaction term $Q$ allows the cosmological constant
($p=-\rho=const$) to become a dynamical quantity, always respecting
the EoS $p(t)=-\rho(t)$. An important feature of this interpretation
is that it gives an approach to obtain analytical solutions. By
selecting a suitable $Q$ one easily integrates the equations for FRW
scalar field cosmologies in closed form. We explicitly obtain exact
solutions in cosmological time $t$, and the potentials can be
written as functions of the scalar field $\phi$, provided that the
scalar field and its potential are invertible functions. As we have
illustrated in previous sections many solutions sparsely found in
the literature result from specific considered choices of the
interacting term $Q$. On the other hand, it is useful to note that
Eqs.~(\ref{CE1A}) and~(\ref{CE2A}) imply that we can determine the
form of the interacting term $Q$ for a given scalar field $\phi$
with its potential $V$.

From the discussed interpretation we have that cosmological models
used for describing the present accelerated expansion of the
Universe and filled with a quintessence, in the form of a scalar
field, and a dark matter component may be interpreted as cosmologies
filled with three fluids: a dust component, a stiff perfect fluid
and an ``effective vacuum energy". In this case we can have models
where the dark matter component does not interact with any other
component, satisfying the standard conservation equation, while
other two components interact with each other, or scenarios where
all three fluid matter components interact with each other.

Since the behavior of any scalar field cosmology can be emulated by
a mixture of two interacting perfect fluids with equation of state
$\omega_{1,2}=\pm 1$, it becomes clear that the behavior of perfect
fluid interacting cosmologies is richer than the one of scalar field
cosmologies, since it can be considered also state parameter values
$\omega_{1,2} \neq \pm 1$. This is more clear for cosmologies which
crosses the phantom divide. In general, the crossing of the phantom
divide cannot be achieved with a unique scalar field. The simplest
model is constructed with the help of two scalar fields, namely one
canonical and one phantom, and the DE is attributed to their
combination~\cite{Cai}. However, such a crossing can be done with
the help of two interacting~\cite{Zhang15} or
non-interacting~\cite{Nesseris} perfect fluids with state parameters
$\omega_1<-1$ and $\omega_2>-1$.

Lastly, let us note that there are in the literature various studies
dealing with relations between fluids and scalar fields. For example
in Ref.~\cite{Bamba15} such relations are studied in the framework
of dark energy cosmological models. However, the proposed procedure
gives a way to express a fluid model as an explicit single or
multiple scalar field theories. In some sense, this is the opposite
to the interpretation of a scalar field discussed in this paper.

In Refs.~\cite{Nojiri2011,Nojiri2007,Nojiri2006,Capozziello2006}
also are studied relations between fluids and scalar fields.
Nevertheless, in these papers it is developed a reconstruction
program for scalar field cosmologies in Einstein theory and in the
framework of modified gravities such as scalar-tensor theory, f(R),
F(G) and string-inspired, scalar-Gauss-Bonnet gravity. In this
scheme, the authors use a generating function $f$ in order to find
cosmological models with one or more scalar fields. It is
interesting to note that in the mentioned references the
gravity-scalar system contains a potential and a scalar coupling
function in front of kinetic term. Our equivalent presentation for
scalar fields, stating that a scalar field in FRW cosmology may be
interpreted as a mixture of an interacting stiff fluid with an
effective vacuum one, can be applied in particular to the scalar
field solutions discussed in
Refs.~\cite{Nojiri2011,Nojiri2007,Nojiri2006,Capozziello2006}. This
is possible since the authors do not separate a scalar field source
into two interacting perfect fluids, as we proposed. This work is
currently in progress and should be available in the near future.

\section{Acknowledgements}
This work was supported by CONICYT through Grants FONDECYT N$^0$
1080530 (MC), N$^0$ 21070949 (FA) and N$^0$ 21070462 (PM)  and by
Direcci\'on de Investigaci\'on de la Universidad del B\'\i o--B\'\i
o through grants N$^0$ DIUBB 121007 2/R and N$^0$ GI121407/VBC (MC,
FA, PM).

\bibliography{myrefs}

\begin{thebibliography}{}

\bibitem{Alabidi} Alabidi, L. and Lyth, D.~H.: JCAP {\bf 0605}, 016 (2006)
(2010)

\bibitem{Felippe8} Avelino, A. and Nucamendi, U.:
  JCAP {\bf 1008}, 009 (2010)

\bibitem{Felippe11} Avelino, A. and ~Nucamendi, U.:
  JCAP {\bf 0904}, 006 (2009)

\bibitem{Bamba15} Bamba, K. ~Capozziello, S., ~Nojiri S.~'i. and
~Odintsov, S.~D.:
  Astrophys.\ Space Sci.\  {\bf 342}, 155 (2012)

\bibitem{Felippe15} Bamba, K., Geng, C.Q., Nojiri, S.'i. and
~Odintsov, S.D.:
  Phys.\ Rev.\ D {\bf 79}, 083014 (2009)

\bibitem{Felippe2} Banijamali, A. and
Fazlpour, B.:
  JCAP {\bf 1201}, 039 (2012)

\bibitem{Felippe5} Banijamali, A. and
Fazlpour, B.:
  Phys.\ Lett.\ B {\bf 703}, 366 (2011)

\bibitem{Felippe4} Brevik, I., Elizalde, E., Nojiri, S.~'i. and
Odintsov, S.D.:
  Phys.\ Rev.\ D {\bf 84}, 103508 (2011)

\bibitem{Felippe6} Brevik, I., Gorbunova, O. and
Saez-Gomez, D.:
  Gen.\ Rel.\ Grav.\  {\bf 42}, 1513 (2010)

\bibitem{Cai} Cai, Y.F., Saridakis, E.N., Setare, .R. and
Xia, J.Q.: Phys.\ Rept.\  {\bf 493}, 1 (2010)

\bibitem{Felippe16} Cai, Y.F. and Wang, J.:
  Class.\ Quant.\ Grav.\  {\bf 25}, 165014 (2008)

\bibitem{Copeland8} Capozziello, S., Piedipalumbo, E., Rubano, C.
and P.~Scudellaro:
  Phys.\ Rev.\ D {\bf 80}, 104030 (2009)

\bibitem{Capozziello2006} Capozziello, S., ~Nojiri, S.~'i. and
~Odintsov, S.~D:
  Phys.\ Lett.\ B {\bf 632}, 597 (2006)

\bibitem{Linde} Cardenas., V.~H.:
  Phys.\ Rev.\ D {\bf 75}, 083512 (2007)

\bibitem{Linde1} Cardenas., V.~H.:
  Phys.\ Rev.\ D {\bf 73}, 103512 (2006)

\bibitem{Cataldo915} Cataldo, M, ~Mella, P., ~Minning, P. and
~Saavedra, J.:
  Phys.\ Lett.\ B {\bf 662}, 314 (2008)


\bibitem{Copeland11} Cataldo, M. and ~del Campo,
S.:
  Phys.\ Rev.\  D {\bf 62}, 023501 (2000)

\bibitem{Copeland1}  Charters, T. and ~Mimoso,
J.~P.:
  JCAP {\bf 1008}, 022 (2010)

\bibitem{Copeland7} Charters, T. and ~Mimoso, J.~P.: JCAP
{\bf 1008}, 022 (2010)

\bibitem{Felippe3} Chattopadhyay, S.:
  The European Physical Journal Plus {\bf 126}, 130 (2011)

\bibitem{Felippe7} Chattopadhyay, S. and ~Debnath,
U.:
  Can.\ J.\ Phys.\  {\bf 88}, 933 (2010)

\bibitem{Chattopadhyay} Chattopadhyay, S. and ~Debnath, U.: Astrophys.\ Space Sci.\  {\bf 326}, 155
(2010)

\bibitem{Chimento15A}  Chimento, L.: Phys. Lett. B {\bf 633}, 9 (2006)

\bibitem{Chimento15}  Chimento, L.: Phys. Rev. D {\bf 65}, 063517
(2002)

\bibitem{Copeland} Faraoni, V.:
  Phys.\ Lett.\ B {\bf 703}, 228 (2011)

\bibitem{Copeland3} Copeland, E.~J., ~Sami, M. and ~Tsujikawa, S.: Int.\ J.\ Mod.\ Phys.\  D
{\bf 15}, 1753 (2006)

\bibitem{Felippe17} Cruz, N., ~Lepe, S. and ~Pena,
F.:
  Phys.\ Lett.\ B {\bf 646}, 177 (2007)

\bibitem{Copeland2} D'Ambroise, J.:
  arXiv:1005.1410 [gr-qc]

\bibitem{Elizalde} Elizalde, E., ~Nojiri, S.~'i. and
~Odintsov, S.~D.:
  Phys.\ Rev.\ D {\bf 70}, 043539 (2004)

\bibitem{Chattopadhyay1} Farajollahi, H., ~Mohamadi, N. and ~Amiri, H.: Mod.\ Phys.\
Lett.\ A {\bf 25}, 2579 (2010)

\bibitem{Felippe10} Felippe, L., Rodrigues, S. and ~Opher, R.: Phys.\ Rev.\  D {\bf 82},
  023501 (2010)

\bibitem{Felippe14} Felippe, L., ~Sun, C.B., ~Wang, J.~-L. and ~Li,
X.~-Z.:
  Int.\ J.\ Mod.\ Phys.\ D {\bf 18}, 1303 (2009)

\bibitem{Felippe} Ghosh, R., ~Chattopadhyay, S. and ~Debnath,
U.:
  Int.\ J.\ Theor.\ Phys.\  {\bf 51}, 589 (2012)

\bibitem{Copeland9} Giovannini, M.:
  Int.\ J.\ Mod.\ Phys.\ A {\bf 22}, 2697 (2007)

\bibitem{Hinshaw} Hinshaw, G. {\it et al.}  [WMAP Collaboration]: Astrophys.\ J.\ Suppl.\
{\bf 180}, 225 (2009)

\bibitem{Felippe13} Ito, Y. and ~Nojiri, S.~'i.:
  Phys.\ Rev.\ D {\bf 79}, 103008 (2009)

\bibitem{Felippe9} Jamil, M. and ~Farooq, M.~U.:
  Int.\ J.\ Theor.\ Phys.\  {\bf 49}, 42 (2010)

\bibitem{Felippe12} Jamil, M., ~Farooq, M.~U. and ~Rashid,
M.~A.:
  Eur.\ Phys.\ J.\ C {\bf 61}, 471 (2009)

\bibitem{Chattopadhyay5} Kofinas, G., ~Panotopoulos, G. and
~Tomaras.T.~N.: JHEP {\bf 0601}, 107 (2006)

\bibitem{Linde5} Kofman.L., ~Linde, A.~D. and ~Starobinsky,
A.~A.:
  Phys.\ Rev.\ Lett.\  {\bf 73}, 3195 (1994)

\bibitem{Linde4} Linde, A.~D.:
  Phys.\ Rev.\ D {\bf 49}, 748 (1994)

\bibitem{Linde6} Linde, A.~D.:
  Phys.\ Lett.\ B {\bf 129}, 177 (1983)

\bibitem{Nesseris} Nesseris, S. and ~Perivolaropoulos,
L.:
JCAP {\bf 0701}, 018 (2007)

\bibitem{Nojiri2011} Nojiri, V. and ~Odintsov,
S.~D.:
  Phys.\ Rept.\  {\bf 505}, 59 (2011)

\bibitem{Nojiri2007} Nojiri, S.~'i. and ~Odintsov,
S.~D.:
  J.\ Phys.\ Conf.\ Ser.\  {\bf 66}, 012005 (2007)

\bibitem{Chattopadhyay4} Nojiri, S.~'i. and ~Odintsov, S.~D.: Phys.\ Lett.\  B
{\bf 639}, 144 (2006)

\bibitem{Nojiri2006} Nojiri, S.~'i. and
~Odintsov S.~D.:
  Gen.\ Rel.\ Grav.\  {\bf 38}, 1285 (2006)


\bibitem{Linde2} Nojiri, S.~'i. and ~Odintsov,
S.~D.:
  Phys.\ Rev.\ D {\bf 68}, 123512 (2003)

\bibitem{Copeland4} Ren, J. and ~Meng, X.~H.: Int.\ J.\ Mod.\
Phys.\  D {\bf 16} 1341, (2007)


\bibitem{Copeland5} Ren, J. and ~Meng, X.~H.: Phys.\
Lett.\ B {\bf 636}, 5 (2006)

\bibitem{Chattopadhyay2} Setare, M.~R. and ~Sheykhi, A.: Int.\
J.\ Mod.\ Phys.\  D {\bf 19}, 1205 (2010)

\bibitem{Chattopadhyay3} Tomaras, T.~N.:
arXiv:hep-ph/0610412

\bibitem{Copeland6} Wang, F. and ~Yang, J.~M.:
  Eur.\ Phys.\ J.\ C {\bf 45}, 815 (2006)

\bibitem{Copeland10} Williams, F.~L., ~Kevrekidis, P.~G.,
  ~Christodoulakis, T., ~Helias, C., ~Papadopoulos, G.~O. and ~Grammenos,
  T.:
  gr-qc/0408056

\bibitem{Zhang15} Zhang, H.:
arXiv:0909.3013 [astro-ph.CO]

\bibitem{Linde3} Zimdahl, W., ~Pavon, D. and ~Maartens,
R.:
  Phys.\ Rev.\ D {\bf 55}, 4681 (1997)



\end{thebibliography}

\end{document}